\documentclass[aps,prl,twocolumn,groupedaddress]{revtex4}
\usepackage{graphicx}

\begin{document}

\title{Ab initio study of mirages and magnetic interactions  
in quantum corrals}
\author{V.S. Stepanyuk $^{(1)(*)}$, L. Niebergall $^{(1)}$, W. 
Hergert $^{(2)}$ and  P. Bruno $^{(1)}$}  

\affiliation{$^{(1)}$ Max-Planck-Institut f\"ur Mikrostrukturphysik,
Weinberg 2, D-06120, Halle, Germany}
\affiliation{$^{(2)}$  Fachbereich Physik, Martin-Luther-Universit\"at,
Von-Seckendorff-Platz 1, D-06120, Halle, Germany}

\date{\today}
\begin{abstract} 

The state of the art ab initio calculations of quantum mirages, 
the spin-polarization of surface-state electrons 
and the exchange interaction between 
magnetic adatoms in  Cu and Co corrals on Cu(111) are presented.
We find that the spin-polarization of the surface-state electrons 
caused by 
magnetic adatoms can be projected  to a remote location and 
can be strongly 
enhanced in  corrals compared to an open surface.
Our studies give a clear evidence that  quantum corrals  
could permit to tailor the  
 exchange interaction between magnetic adatoms at large separations.  
The spin-polarization of surface-state electrons at the empty focus     
in the Co corral used in the experimental setup of Manoharan et 
al., (Nature {\bf 403}, 512 (2000)) is revealed.

\end{abstract}
\maketitle

As the physical size of a system approaches atomic dimensions, quantum 
effects are known to play a significant role.  One of the most   
striking illustrations of the quantum behavior in atomic-scale 
nanostructures  is  
the observation of the electronic confinement of surface-state electrons 
in the Fe corral constructed  in an atom-by-atom fashion on a copper (111)
surface \cite{c1}.    
The structures which  confine the electrons on surfaces
can be built using the manipulation into the required geometry  of
individual adsorbed atoms by the scanning tunneling 
microscope (STM) \cite{c2}. 
Altering the size and shape of 
artificial structures, one could  affect their quantum states. 
The controllable modification of quantum states could permit to 
manipulate individual  spins, their dynamics, interactions, and 
could be of a great importance for the development of quantum 
nano-devices.

Recent remarkable experiments of Manoharan et al., \cite{c3} have shown 
that the 
electronic structure of adatoms can be projected to a remote location
exploiting  quantum confinement of electronic states in an 
engineered nanostructures.  
Placing an  atom of magnetic cobalt at one focus of the   
elliptical corral constructed  from several dozen cobalt atoms on Cu(111)
caused the Kondo mirage to appear at the empty focus.  Results of this 
fascinating experiment have been explained by Fiete et al., \cite{c4} 
using the 
scattering 
theory. They have demonstrated that the mirage at the empty 
focus of the 
elliptical corral is the result of resonant scattering of electrons from 
the magnetic adatom and scattering from the adatoms of the walls of the 
corral. There have also been several important 
theoretical studies related to quantum corrals and 
the mirage experiments [5-9].

Although above mentioned works  have provided an appealing picture
of quantum mirages and  interactions in quantum corrals, a full 
understanding of the behavior of surface-state electrons in man-made 
nanostructures and   their response to magnetic adatoms 
requires a first-principle calculations. Such studies would be of 
fundamental interest for our understanding of magnetism at the atomic scale, 
and could establish a basis for 
the controllable manipulation of quantum states    
in atomic-scale nanostructures.
Obviously, the challenge  would be to tailor  the interaction 
between single spins.    

In this Letter we present a fully ab initio study of  quantum 
mirages, 
the spin-polarization and the exchange interaction between magnetic 
adatoms in 
corrals. We concentrate on 3d adatoms in elliptical Cu 
and Co corrals on Cu(111).  
We demonstrate  that the interaction of 
magnetic adatoms 
with the confined surface-state electrons of corrals  leads to 
significant changes  in electronic and magnetic states of corrals,  
and could produce a mirage at a remote location.
We show that the spin-polarization of surface-state 
electrons caused 
by magnetic adatoms placed in the corral focus is projected  
to an empty 
focus. 
Strong enhancement of  the spin-polarization 
in corrals compared to an open Cu(111) surface is found.    
Our study  presents a clear evidence that
the long-range exchange interaction between magnetic adatoms is
strongly affected by confined electronic  states of corrals. The 
possibility of 
tailoring the exchange interaction by modifying  the corral geometry 
is demonstrated. 
The spin-polarization of the electron gas in the empty focus 
of the Co corral 
used in the experimental 
setup of Manoharan et al. \cite{c3} is revealed.

Adatoms and corrals destroy the two-dimensional (2D) periodicity of the 
ideal 
surface. Heller et al. \cite{c10} have shown 
in their studies of 'quantum stadium' 
 that 
the multiple-scattering approach is physically motivated 
to treat  the electronic states of an arbitrary 
corral geometry and arbitrary placed
adatoms in  2D  systems. Therefore, we believe that an ab 
initio method based on the multiple scattering theory 
is  well suited 
for calculations of magnetic adatoms in quantum corrals.
Our approach is based 
on the density functional theory (DFT) and
multiple-scattering approach  using the Korringa-Kohn-Rostoker (KKR)
Green's function method for adatoms and clusters on 
surfaces \cite{c11}. Although  the DFT does not account for 
properties of
dynamical origin like the Kondo effect, it is an accurate
method to determine static quantities \cite{Uj}. Therefore, our
calculations  are related  to electronic and magnetic
properties of quantum corrals above the Kondo
temperature. 

The Green's function of the corral (with or without adatoms) 
is  calculated in a real
space representation.
\begin{eqnarray}
G_{LL'}^{nn'}(E)=
\mbox{\r G}_{LL'}^{nn'}(E)+
\sum\limits_{n''L''}
\mbox{\r G}_{LL''}^{nn''}(E)
\Delta t_{L''}^{n''}(E)
G_{L''L'}^{n''n'}(E),
\end{eqnarray}
where $G_{LL'}^{nn'}(E)$  is the energy-dependent structural
Green's function matrix and $\mbox{\r G}_{LL''}^{nn''}(E)$ the
corresponding matrix for the ideal surface; 
$\Delta t_{L}^{n}(E)$ describes the difference in the
scattering properties at site $n$ induced by the existence of the
corral and adatoms \cite{c12,c13}
The atomic structure of the substrate is taken into account in 
calculations \cite{c14}. 

First, we consider the elliptical Cu corral on Cu(111) (Fig.1). 
\begin{figure}[ht]
\vspace*{0.5cm}
\includegraphics[width=7.0cm]{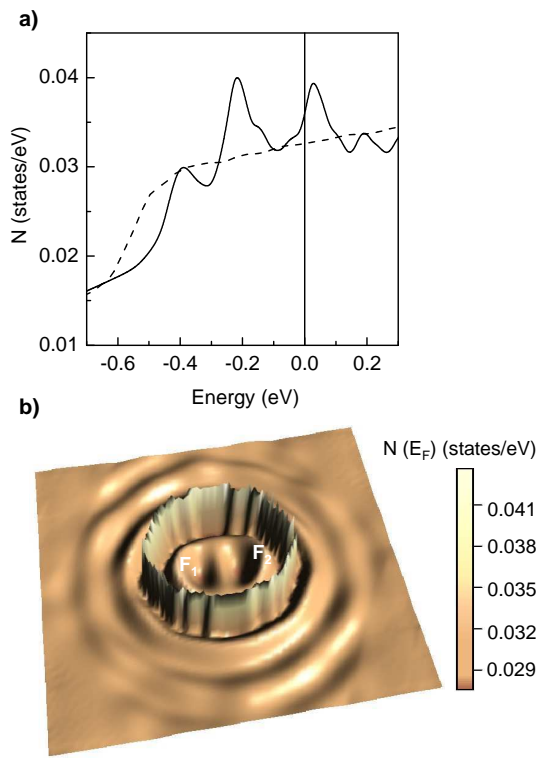}
\label{fig1}
\caption{(color) a: The LDOS at the corral focus; b: Quantum interference 
patterns inside and outside of the corral. The elliptical Cu corral with 
semi-axis a= 25 \AA, and eccentricity $\varepsilon =0.5$ on Cu(111) is 
presented; the distance between nearest Cu atoms in the corral walls  
is equal to the nearest neighbor separation on the Cu(111) surface. The 
LDOS of an open Cu(111) surface is shown by the dashed line}
\end{figure}
The quantum interference between the electron waves traveling
towards the  Cu atoms forming the  corral wall and  
the backscattered ones leads to the confinement of the surface-state 
electrons inside the corral \cite{c15}. 
The energy resolved local density of states (LDOS) at  one of the 
corral's foci 
is presented in Fig.1a. It is  seen that the LDOS
exhibits a series of resonant peaks indicating the quantum confinement.
The spatial distribution of the LDOS at the Fermi energy   
is presented in Fig1b. The standing wave 
patterns outside and inside the corral are  caused by the quantum 
interference of surface-state electrons scattered by atoms of the corral 
wall. 
The oscillations of  the LDOS outside the corral at  distances larger 
than $8~\AA$ are well described   by the period of 
about $15~\AA$ which is close to  
half of the Fermi wavelength of  surface-states electrons on 
Cu(111) \cite{c16}. 

If a magnetic adatom is placed at the  focus of the corral, the 
resonance scattering of the surface-state  electrons by  the adatom 
and the corral walls leads 
to  striking changes in the LDOS. As an example, we show in Fig.2 our 
calculations for Co adatom. Strong changes in the LDOS near 
$E_F$ in the empty focus are resolved (see Fig.2a.). 
\begin{figure}[ht]
\vspace*{0.4cm}
\includegraphics[width=7.0cm]{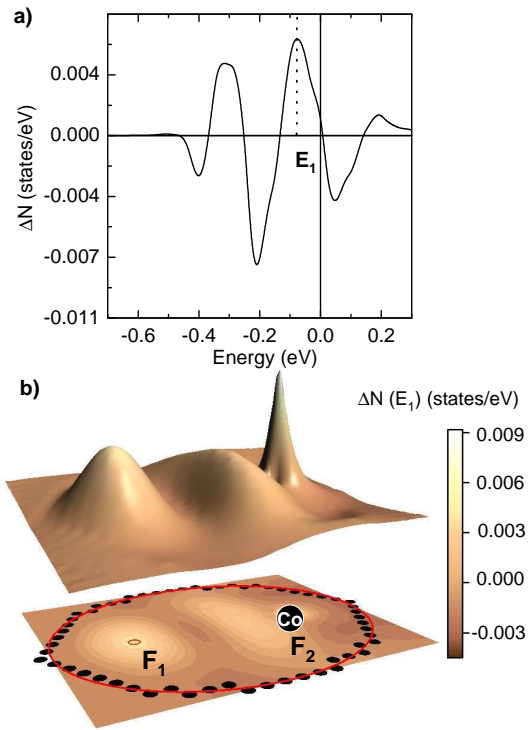}
\label{fig2}
\caption{(color) (a) The magnetic Co adatom is located in the 
Cu corral focus.
The LDOS in the empty corral focus is depicted; (b) The mirage effect in 
the corral:  
the changes in the LDOS inside the corral with the magnetic adatom 
with  respect to  the empty Cu corral are illustrated; the LDOS 
of the single Co adatom on an open
Cu(111) surface has been subtracted from the image.}   
\end{figure}
Comparing this 
LDOS with the one for empty corral (Fig.1a), it is 
evidently that  
resonances act as waveguides for the projection of the electronic 
structure of the magnetic adatom to an empty focus \cite{c3}.
The change in the LDOS (close to the $E_F$) at the  
empty focus clearly 
demonstrates the mirage effect (Fig.2b).
To the best of our knowledge, the above  result is the first fully ab 
initio  confirmation of  
the projection of electronic structure of the magnetic adatom to a 
remote location. 

Another very important consequence of the quantum confinement in the 
corral   concerns  the spin-polarization of the 2D electron 
gas. We place the magnetic Co adatom at the  focus of the Cu corral 
and 
calculate  the energy-resolved spin-polarization at the empty  
focus. 
Results shown in Fig.3a  reveal a strong enhancement of the 
spin-polarization in the empty focus of the corral compared to that around  
Co adatom on an open Cu(111) surface. 
\begin{figure}[ht]
\vspace*{0.4cm}
\includegraphics[width=7.0cm]{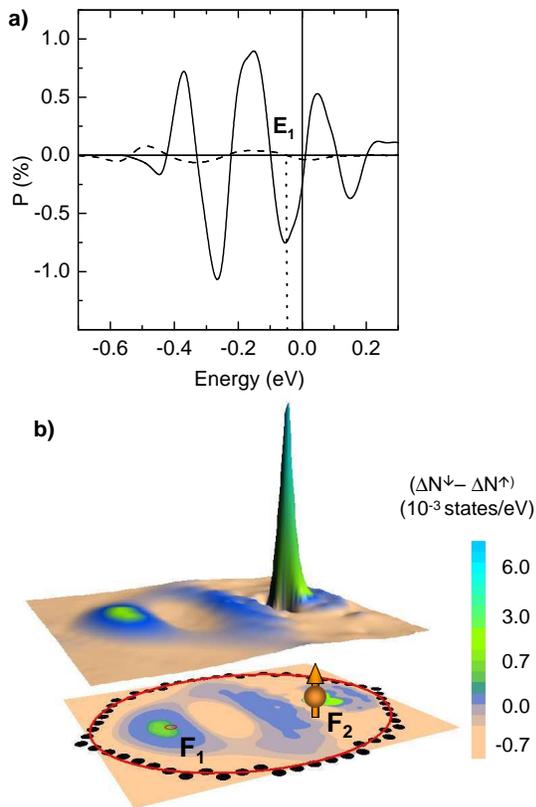}
\label{fig3}
\caption{(color) Enhancement of the spin-polarization at  the  
empty  
focus of the Cu corral. The magnetic Co adatom is placed at  the right 
focus. a: The spin-polarization  is determined as: 
($N\uparrow$ - 
$N\downarrow$)/($N\uparrow$ 
+$N\downarrow$), where $N\uparrow$ and  $N\downarrow$) are 
the LDOS 
for majority and 
minority electrons 
respectively. The spin-polarization around the Co adatom 
on the  open Cu(111) surface is shown by dashed line; b: 
$\Delta N\downarrow$ and 
$\Delta N\uparrow$ are determined by the difference between 
LDOS  at the 
energy $E_1$ between the corral with the Co adatom and the  
single Co adatom on the open Cu(111).}
\end{figure}
Our calculations clearly demonstrate 
that the spin-polarization of surface-state electrons at the empty focus 
is  very close to that near the magnetic adatom (Fig.2b).
In other words, the spin-polarization is projected to the second focus by 
the quantum states of the corral.
We emphasize  that the corral walls are non-magnetic and 
therefore, the 
spin-polarization at the second focus  is only      
caused by the spin-dependent scattering of the  
surface-state electrons by the magnetic adatom.
This result is the example of a possible   
'magnetic information transfer' on metal surfaces. 
Recently, a theory of magnetic mirages based on the Anderson model has 
been proposed by Hallberg et al. \cite{c6}.
Our ab initio studies  are  in line with this theory and unambiguously 
prove that  tailoring  the 
spin-polarization of 2D electron gas could be achieved in artificial 
atomic structures by exploiting the quantum confinement of surface-state 
electrons.

The above results suggest that there could be a 
significant impact of the 
quantum confinement in corrals on the interaction between magnetic 
adatoms.  

To give a clear evidence that quantum corrals  can be used for controlled 
modification of magnetic interactions, we perform ab initio calculations 
for the exchange interaction between 3d adatoms inside  the Cu corral. 
For large interatomic separations  the exchange interaction energies are 
very small (meV and  $\mu$eV). 
Therefore, there is the problem of subtracting huge total 
energy values to obtain the resulting small interaction energies. 
However, it has been proved that applying the force theorem and using  
the single-particle energies, instead of total energies, one can resolve   
very small  interaction energies  
at large atomic distances with  high accuracy \cite{c17}. 

We have calculated the exchange interaction between 3d adatoms  in 
the Cu 
corral on Cu(111) for different adatom-adatom separations. 
In the absence of the corral, i.e. for an open surface, 
the exchange interaction between magnetic
adatoms is dominated  by the surface-state electrons, and its  
magnitude decays as $1/d^2$ (d is the 
adatom-adatom separation) \cite{c18}.  
However, the quantum corral drastically influences on  the  interaction 
between magnetic adatoms. This is well seen in 
Fig.4 where our calculations  for  
3d adatoms placed in the corral foci are presented.  
\begin{figure}[ht]
\vspace*{0.4cm}
\includegraphics[clip,width=7.0cm]{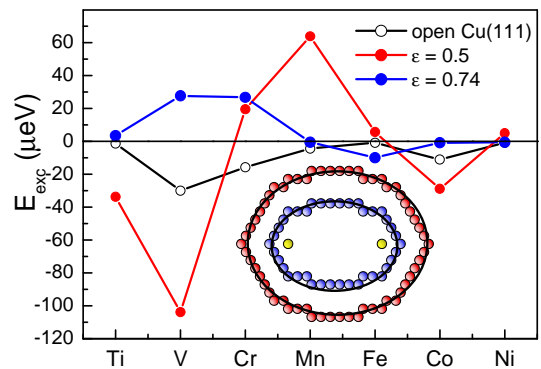}
\label{fig4}
\caption{(color) a: The exchange interaction between magnetic 
adatoms inside the 
Cu 
corrals of different eccentricities;  the distance between foci is fixed, 
The exchange interaction on an open Cu(111) is presented by the black 
line.}  
\end{figure}

In order to demonstrate  the effect of the corral
geometry on the exchange interaction,
we show  our calculations for the Cu corrals of
different eccentricities.
These  striking results  reveal that    
the exchange interaction in the corral is  strongly enhanced compared to 
an open surface, and can  switch  from the ferromagnetic coupling to 
the antiferromagnetic one by modifying the corral geometry. 
We believe that these calculations  provide the clearest evidence of   
tailoring  the magnetic interactions between adatoms 
at large distances by 
 constructing appropriate corrals \cite{c19}.  Note that recent 
results of Lazarovits et al.,
have also indicate that the geometry of corrals may affect
their  electronic and magnetic states \cite{Laz}.

Finally, we  apply our method for calculations of the spin-polarization in 
the Co corral used in the experimental setup  of  Manoharan et 
al., \cite{c3}. 
A net spin-polarization of the electron gas in the Co corral  was 
suggested in this work as the possible reason for  the quantum mirage 
in an empty focus. 
First, we have found  the LDOS for the minority and majority 
electrons 
inside the Co corral with the Co adatom placed in the corral focus. 
Calculations have been performed for energies close to the Fermi 
energy ($E_F$+10meV). 
Then, we have repeated the calculations for an empty Co corral.   
In fact, the difference between the two calculations 
presents the effect of the Co adatom on the spin-polarization of the 
electron gas inside the corral.  However, the spin-polarization in the 
empty focus is found to be significantly smaller than the 
spin-polarization on the Co adatom. Therefore, to make a clear 
presentation of the magnetic mirage, i.e., enhanced spin-polarization at 
the empty focus, the spin-polarization of the single Co adatom on an open 
Cu(111) surface has been removed from the image shown in Fig.5. 
\begin{figure}[ht]
\vspace*{0.4cm}
\includegraphics[clip,width=7.0cm]{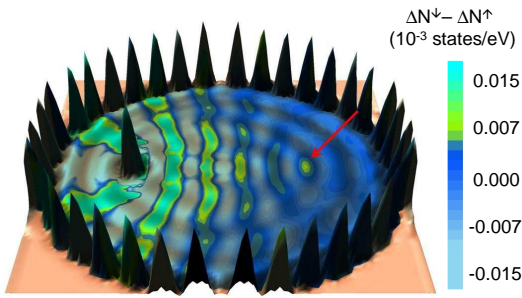}
\label{fig5}
\caption{(color) The LDOS at
the Fermi energy on the Co adatom and the Co
atoms of the corral walls are shown. 
The spin-polarization of 
surface-state 
electrons inside  the Co corral is presented in color: 
$\Delta N\downarrow$ and
$\Delta N\uparrow$ are determined  by the difference between
LDOS near the Fermi energy (+10 meV)
of  the Co corral with the Co adatom, the empty 
Co corral  and the
single Co adatom on the open Cu(111). Mirage in the 
empty focus is marked by the red arrow.
The geometrical parameters of the corral
are the same as in  the experimental setup of ref.3, i.e.,
semi-axis a= 71.3 \AA, and eccentricity $\varepsilon =0.5$.}
\end{figure}
The oscillations  of the spin-polarization  
are well seen in Fig.5. The enhancement  of the  
spin-polarization in an empty focus is revealed.

In summary, we have presented the first ab initio studies of quantum 
mirages and the magnetic interactions in quantum corrals. While we have 
used a 
particular systems, Cu and Co corrals on Cu(111) to illustrate several 
effects of the quantum confinement of surface-state electrons,  the 
main conclusions of our work are independent of the specific systems. It 
is generally true that the spin-polarization of surface electrons caused 
by magnetic adatoms can be projected to a remote location by quantum 
states of corrals, and the exchange interaction between magnetic atoms can 
be manipulated at large distances.
Our work opens up new possibilities  for experimental and 
theoretical 
studies of magnetic properties in engineered atomic nanostructures. 

This work was supported by DFG, Schwerpunktprogramm 'Cluster in 
Kontakt mit Oberfl\"achen: Elektronenstruktur und Magnetismus'.


\begin{references}
\bibitem{c13} Our calculations have proved that only a limited number
of atoms near the corral wall and near the adatoms placed inside the
corral contribute to $\Delta t_{L}^{n}(E)$ in eq.1. Therefore,
even very large corrals can be calculated fully ab initio using
our approach. We have found that  self-consistent and 
 non-self-consistent calculations  give essentially the same results for 
the LDOS, the spin-polarization and  
the interaction energies,  provided that the self-consistent potentials 
of single adatoms on Cu(111) are used in the non-self-consistent 
calculations.   
The effect of bulk electronic states on the scattering of 
the 
surface electrons is taken into account in our approach.
\bibitem{c14}  We have found that atomic relaxations of atoms of
the corral walls and adatoms inside the corral practically do not 
influence
the main results of our work.
\bibitem{c19} We have found that changes of  the eccentricity  
influence  on the position 
of the resonances in the LDOS  close to the $E_F$ in the  
corral.
These resonances form the projection medium for the long-range interaction 
between spins (see also ref.3), therefore  modifying their 
positions near 
the $E_F$ one can tailor   the exchange interaction between 
magnetic adatoms.
\end{references}
\end{document}